\documentclass[twocolumn,pra]{revtex4}

\begin{document}

\title{Differences between the trajectory representation and
Copenhagen \\
regarding the past and present in quantum theory}

\author{Edward R.\ Floyd}
\affiliation{10 Jamaica Village Road, Coronado, CA 92118-3208,
USA}
\email{floyd@mailaps.org}

\date{29 May 2003}

\begin{abstract}
We examine certain pasts and presents in the classically forbidden
region.  We show that for a given past the trajectory
representation does not permit some presents while Copenhagen
predicts a finite probability for these presents to exist.  This
suggests another gedanken experiment to invalidate either
Copenhagen or the trajectory representation,
\end{abstract}

\maketitle

\section{INTRODUCTION}

While Copenhagen postulates that the wave function $\psi $
contains an exhaustive knowledge of quantum phenomenon, the
trajectory representation (TR) in contrast considers $\psi $ to be
incomplete because an eigenfunction $\psi $ with energy $E$ may
have microstates.$^{\ref{bib:floyd82}-\ref{bib:carroll99}}$
Copenhagen makes its predictions based upon $\psi $ being
complete. Copenhagen for given past conditions predicts only the
probabilities of later events.$^{\ref{bib:bohm}}$  If Copenhagen's
probability density, $\psi ^{\dagger} \psi $, is finite for some
spatial point for a particular present, then that present is
permitted.  We define $\{\mbox{Copenhagen}\}$ to be the domain of
the presents allowed by Copenhagen.  On the other hand, TR is a
causal representation. Each microstate has its own trajectory. The
present for a given past is microstate dependent. In turn, we
define $\{\mbox{TR}\}$ to be the domain of the presents allowed by
TR. The question we investigate here is whether the family of
trajectories of the microstates of a given wave function, $\psi $,
for a given past covers the spatial domain allowed by Copenhagen
at a later time. In other words, is $\{ \mbox{TR}\} \cup \{
\mbox{Copenhagen}\} = \{ \mbox{TR}\}$ always true? It turns out
that sometimes it isn't true. We show where and how for a given
initial (past) postion the family of trajectories of TR does not
cover the domain of present positions allowed by Copenhagen.

Other differences between Copenhagen and TR have previously been
presented. Recently, Copenhagen and TR have been shown to predict
differences regarding perturbing impulses$^{\ref{bib:floyd99}}$
and regarding the overdetermination of the
motion$^{\ref{bib:floyd00}}$ by a redundant set of constants of
the motion.  For a perturbing impulse, TR, which is a causal
theory couched in a Hamilton-Jacobi formulation, has the
perturbing impulse act on a particle at its instantaneous position
while Copenhagen, as a probability theory, mandates that even
impulses must be averaged over Copenhagen's probability density
$\psi ^{\dagger} \psi $.  We note that for adiabatic
perturbations, the trajectory representation and Copenhagen are in
agreement. With regard to the overdetermination of the of the
motion, the generator of the motion$^{\ref{bib:floyd82}}$ and the
subsequent trajectory can be described by a finite number of
constants of the motion.  Any overdetermination of the motion by a
super-sufficient number of constants of the motion would still
have only a prescribed number of independent constants of the
motion while all surplus constants of the motion would be
redundant as they could be described in terms of the independent
constants of the motion.  Meanwhile, Copenhagen regards measuring
these surplus constants of the motion as independent measurements.
This investigation herein needs only a sufficient set of constants
of the motion to determine the microstate to present differences
between Copenhagen and TR.  We note for completeness that position
and momentum are insufficient by themselves to specify motion
regardless of the Heisenberg uncertainty
principle.$^{\ref{bib:fm},\ref{bib:carroll00},\ref{bib:floyd00a}}$

In Section 2, we show that, for a given past, TR does not generate
trajectories that cover the Copenhagen-allowed domain of present
positions when we specify an initial (past) position in the
classically forbidden region inside a semi-infinite step barrier.
In Section 3, we show that for a square well TR does generate
trajectories that cover the Copenhagen-allowed domain of present
positions. We discuss our results in Section 4.  This
investigation borrows the computational results from prior TR
investigations that had examined other aspects of quantum theory.
The interested reader who may be uninitiated in TR will find an
introduction to TR in ``Extended Version of `The Philosophy of the
Trajectory Representation of Quantum Mechanics'\,'',
quant-ph/0009070.

\section{SEMI-INFINITE STEP BARRIER}

Let us first investigate a case where the family of trajectories
of TR does not cover the domain of present positions allowed by
Copenhagen, $\{\mbox{Copenhagen}\}$. We consider a particle has
sub-barrier energy $E$. We also assume a barrier given by

\[
V=\left\{ \begin{array}{lc}
            0, & \ x < 0 \\
            U > E, & \ x \ge 0
            \end{array}
    \right.
\]

\noindent For this potential, the wave function, $\psi $, is
finite over the finite $({\bf r},t)$-manifold.  We investigate
whether a particle initially posited in a classically forbidden
region of a semi-infinite step barrier can have a family of
microstates whose trajectories span the finite $({\bf
r},t)$-manifold consistent with Copenhagen. Let us now consider
the dwell time, $t_{\mbox{\scriptsize D}}$, that a particle spends
in the a classically forbidden region inside the barrier. This
$t_{\mbox{\scriptsize D}}$ is the time for a particle to complete
its round trip from the wall of the step barrier at $x=0$ out to
its turning point at $x= \infty $ and then its return back to
$x=0$.  This dwell time, $t_{\mbox{\scriptsize D}}$, has been
studied for TR and is given by$^{\ref{bib:floyd00}}$

\begin{equation}
t_{\mbox{\scriptsize D}} = 2 \frac{(ab-c^2/4)^{1/2}[1+(\kappa
/k)^2]}{a \pm c(\kappa /k) + b(\kappa /k)^2} \frac{m}{\hbar \kappa
k} \label{eq:dwelltime}
\end{equation}

\noindent where $k=(2mE)^{1/2}/\hbar$ and $\kappa =
[2m(U-E)]^{1/2}/\hbar $. The coefficients $a,b,c$ are additional
constants of the motion that specify the microstate. One of these
coefficients may be eliminated by noting that TR scales the
Wronskian for the pair of independent solutions of the associated
Schr\"{o}dinger equation to be equal to $\{ 2m/[\hbar
^2(ab-c^2/4)]\} ^{1/2}.^{\ref{bib:floyd00}}$ Two additional
constants of the motion are needed because the quantum stationary
Hamilton-Jacobi equation in one dimension,

\[
\underbrace{\frac{W_x^2}{2m}+V-E}_{\mbox{\footnotesize classical
HJE}}=\underbrace{\frac{\hbar^2}{4m}\left[ \frac{W_{xxx}}{W_x} -
\frac{3}{2} \left(
\frac{W_{xx}}{W_x}\right)^2\right]}_{\mbox{\footnotesize quantum
effects $\Rightarrow $ third-order ODE}}
\]

\noindent is a third-order nonlinear differential equation while
the classical stationary Hamilton-Jacobi equation is only
first-order. By Eq.\ (\ref{eq:dwelltime}), the dwell time is a
function of the microstate coefficients $(a,b,c)$.

Let us now briefly digress on coefficients.  While the particular
values for the coefficients $(a,b,c)$ are dependent upon which set
of independent solutions of the Schr\"odinger equation is chosen
consistent with the normalization of the Wronskian, the observable
results are independent of the particular
choice.$^{\ref{bib:floyd82},\ref{bib:fm}}$ This is analogous to
vector analysis where the results of vector operations remain
independent of the choice of coordinate systems.

Let us now compare our $t_{\mbox{\scriptsize D}}$ calculated by TR
to the barrier dwell times calculated by other workers not using
TR. We consider a monochromatic incident particle ($x<0$). As
shown elsewhere,$^{\ref{bib:floyd00}}$ the particle is
monochromatic iff $a=b$ and $c=0$. Then $t_{\mbox{\scriptsize D}}$
reduces to$^{\ref{bib:floyd00}}$

\[
t_{\mbox{\scriptsize D}}  = 2m/(\hbar \kappa k) = \hbar
/[E(U-E)]^{1/2}
\]

\noindent which is consistent$^{\ref{bib:floyd00}}$ with the dwell times for barriers
presented by Hartman$^{\ref{bib:hartman}}$ and Fletcher$^{\ref{bib:fletcher}}$.

We also note that $t_{\mbox{\scriptsize D}}$ is inversely
proportional to $\kappa $ or $(U- E)^{1/2}$.  Particle velocity
inside the barrier increases as $(U-E)^{1/2}$ increases consistent
with Barton.$^{\ref{bib:barton}}$ This seems to be
counterintuitive and to support superluminal velocities.  As the
trajectory transverses an infinite distance between the interface
at $x=0$ and the turning point at $x=\infty $ in a finite duration
of time, the velocity along the trajectory in this
non-relativistic opus must become infinite for at least an
infinitesimal duration.$^{\ref{bib:floyd00}}$  This is another
manifestation that TR is a nonlocal theory.  Others have made much
ado about this aspect of nonlocality in
tunnelling.$^{\ref{bib:sciam}}$

We now take another brief divergence.  For completeness, we note
that the superluminal velocity of TR implies the particle spends
less time in these regions. Copenhagen alleges that this ``reduce
amount of time in a region'' just manifests a reduced probability
density in this very region.  Here Copenhagen has tacitly assumed
that quantum systems are ergodic where distributions over time
converge to probability densities. In higher dimensions, the
infinite velocity that occurs at the turning point at $\infty$
implies that the trajectory for oblique incidence on the step
barrier is tangentially embedded in the face of the surfaces of
constant reduced action (Hamilton's characteristic function) at
the turning point at infinity.$^{\ref{bib:floyd00}}$ Furthermore,
the trajectory for oblique incidence has a cusp at this turning
point at $\infty$.

We next determine the longest period of time, i.e. the maximum
dwell time, that a particle can remain inside the barrier (we note
that this maximum dwell time represents the dwell time for a
particle whose initial position is on the interface; any other
initial position within the barrier would induce the particle to
remain in the barrier for even less time). For mathematical
convenience, we scale the Wronskian so that $(ab-c^2/4)^{-1/2}=1$.
The maximum dwell time, $t_{\mbox{\scriptsize MD}}$, for a
particle in the step barrier occurs at

\begin{eqnarray*}
a & = & (1+c^2/4)^{1/2}\kappa/k=2^{1/2}\kappa /k, \\
b & = & (1+c^2/4)a^{-1}=2^{1/2}k/\kappa, \\
c & = & 2-\epsilon \ \mbox{where}\ 0<\epsilon \ll 1.
\end{eqnarray*}

\noindent Then the dwell time has a least upper bound given by

\[
t_{\mbox{\scriptsize MD}} < \frac{1+(\kappa /k)^2}{(2^{1/2}-1)}
\frac{m}{\hbar \kappa ^2}
\]

\noindent Hence, if the particle is in the barrier at a particular
time, $t_0$, then, after the time $t_0+t_{\mbox{\scriptsize MD}}$
at the most, the particle must have been reflected from the
barrier and can no longer be in the barrier.  This differs with
Copenhagen, which postulates that the particle could be found
again in the barrier forever. And we conclude that

\[
\{ \mbox{TR}\}_{\mbox{\scriptsize SB}} \cup
\{\mbox{Copenhagen}\}_{\mbox{\scriptsize SB}} \ne
\{\mbox{TR}\}_{\mbox{\scriptsize SB}}
\]

\noindent for the particles reflected from a semi-infinite step
barrier (the subscript ``SB'' denotes step barrier).

\section{SQUARE WELL}

Let us now investigate a case where the family of trajectories of
TR does cover the $({\bf r},t)$-domain allowed by Copenhagen.  We
assume a square well given by

\[
V= \left\{ \begin{array}{lc}
               U, &  \ |x| \ge q \\
               0,      &  \ |x|< q.
           \end{array}
     \right.
\]

\noindent The trajectories for the bound states, $E<U$, for this
square well have already been studied.$^{\ref{bib:floyd00}}$

The period of libration, $t_{\mbox{\scriptsize L}}$, for the
trajectory for microstate $(a,b,c)$ has been determined to
be$^{\ref{bib:floyd00}}$

\begin{eqnarray}
t_{\mbox{\scriptsize{L}}} & = & 4[1+(\kappa /k)^2]
\frac{m(q+\kappa
^{-1})}{\hbar k} \nonumber \\
& & \ \times \frac{(ab-c^2/4)^{1/2} [a+b(\kappa /k)^2]} {a^2 +
(2ab-c^2)(\kappa /k)^2 + b^2(\kappa /k)^4} . \label{eq:libration}
\end{eqnarray}

\noindent Note that Eq. ({\ref{eq:libration}) is valid for
symmetric and antisymmetric microstates of bound
states.$^{\ref{bib:floyd82}}$

The maximum time, $t_{\mbox{\scriptsize ML}}$, for libration is
given at

\begin{eqnarray*}
a & = & (1+c^2/4)^{1/2}\kappa/k=2^{1/2}\kappa /k, \\
b & = & (1+c^2/4)a^{-1}=2^{1/2}k/\kappa, \\
|c| & = & 2-\epsilon \ \mbox{where}\ 0<\epsilon \ll 1
\end{eqnarray*}

\noindent where again the Wronskian has been normalized so that
$(ab-c^2/4)^{-1/2}=1$. We note that here we need the magnitude of
coefficient $c$ for $t_{\mbox{\scriptsize ML}}$ rather than just
its value as we did for $t_{\mbox{\scriptsize MD}}$. This is
because we must consider reflection from both step barriers of the
square well with interfaces at $x=\pm q$. From Eq.\
(\ref{eq:libration}), the maximum time for libration has a least
upper bound given by

\[
t_{\mbox{\scriptsize ML}} < 2^{3/2} [1-(\kappa /k)^2]
\frac{m(q+\kappa ^{-1})}{\hbar \kappa }.
\]

On the other hand, the minimum time for libration has a greatest
lower bound given by zero.  Either coefficient, $a$ or $b$, in Eq.
(\ref{eq:libration}) can grow without finite bound inducing
$t_{\mbox{\scriptsize{L}}}$ to become zero since
\[
t_{\mbox{\scriptsize L}} \propto \frac{\displaystyle
(ab-c^2/4)^{1/2}[a+b(\kappa /k)^2]}{\displaystyle a^2 +
(2ab-c^2)(\kappa /k)^2 + b^2(\kappa /k)^4}
\]

\noindent while still maintaining $(ab-c^2/4)=1$.  Therefore,
there always exists for a bound-state particle a trajectory with
an appropriate $t_{\mbox{\scriptsize L}}$ to connect any past
position with any present. Additional libration periods may be
added consistently to a corresponding sub-orbital duration to
cover $\Delta t
> t_{\mbox{\scriptsize ML}}$.  We find that

\[
\{ \mbox{TR}\}_{\mbox{\scriptsize SW}} \cup\{
\mbox{Copenhagen}\}_{\mbox{\scriptsize SW}} = \{
\mbox{TR}\}_{\mbox{\scriptsize SW}}
\]

\noindent for bound state particles in a square well (the
subscript ``SW'' denotes square well). For bound states of a
square well, Copenhagen's independence of the present from the
past corresponds to the spanning of the $({\bf r},t)$-manifold by
the trajectories for a family of microstates.

For completeness, we note that the converse of the above is not
true for excited states of the square well since

\[
\{ \mbox{TR}\}_{\mbox{\scriptsize SW}} \cup
\{\mbox{Copenhagen}\}_{\mbox{\scriptsize SW}} \ne
\{\mbox{Copenhagen}\}_{\mbox{\scriptsize SW}}
\]

\noindent for excited states.  The reason is that the isolated
zeros in the classically allowed region of the excited wave
function are accessible in TR.$^{\ref{bib:floyd82}}$ Nevertheless,
we note for completeness that every neighborhood of these isolated
zeros of the excited wave function contains points that are
accessible to the corresponding $\{\mbox{TR}\}_{\mbox{\scriptsize
SW}}$.  The isolated zeros of excited states are limit points of
the corresponding $\{\mbox{TR}\}_{\mbox{\scriptsize SW}} \cup
\{\mbox{Copenhagen}\}_{\mbox{\scriptsize SW}}$.

\section{DISCUSSION}}

The semi-infinite step barrier is a counterexample where some
presents are excluded by a single past. Our dwell times for the
particle in the step barrier are shown herein to be consistent
with the findings of other workers. In contrast, overdetermination
of the constants of the motion$^{\ref{bib:floyd00}}$ corresponds
to multiple pasts (i.e., $t_0< t_{00}< t_{000} \cdots $)where a
sufficient number of multiple pasts may overdetermine a
microstate.

Still the step barrier only presents a gedanken experiment to
distinguish TR from Copenhagen because an ideal semi-infinite step
potential is corrupted by reality.

\bigskip

\begin{acknowledgments}
I thank M.\ Matone, A.\ E.\ Faraggi, R.\ Carroll, and A.\ Bouda
for their stimulating discussions and encouragement.
\end{acknowledgments}

\begin{enumerate}\itemsep -.06in

\item \label{bib:floyd82} E.\ R.\ Floyd, Phys.\ Rev.\ {\bf D 26}, 1339 (1982); {\bf 34},
3246 (1986); Found. Phys. Lett. {\bf 9}, 489 (1996), quant-ph/9707051.

\item \label{bib:fm} A.\ E.\ Faraggi and M.\ Matone, Int.\ J.\
Mod.\ Phys.\ {\bf A 15}, 1869 (2000), hep-th/9809127.

\item \label{bib:carroll99} R.\ Carroll, Can.\ J.\ Phys.\ {\bf 77}, 319 (1999),
quant-ph/9903081.

\item \label{bib:bohm} D.\ Bohm, {\it Quantum Theory} (Prentice
Hall, 1951, Englewood Cliffs) pp. 26--29.

\item \label{bib:floyd99} E.\ R.\ Floyd, Int.\ J.\ Mod.\ Phys.\ {\bf A 14}, 1111 (1999),
quant-ph/9708026.

\item \label{bib:floyd00} E.\ R.\ Floyd, Found.\ Phys.\ Lett.\ {\bf 13} 235 (2000),
quant-ph/9708007

\item \label{bib:carroll00} R.\ Carroll, {\it Quantum Theory,
Deformation and Integrability} (Elsevier, 2000, Amsterdam) pp.
50--56.

\item \label{bib:floyd00a} E.\ R.\ Floyd, Int.\ J.\ Mod. Phys.\ {\bf A 15}, 1363 (2000),
quant-ph/9907092.

\item \label{bib:hartman} T.\ E.\ Hartman, J.\ App.\ Phys.\ {\bf
33}, 3247 (1962).

\item \label{bib:fletcher} J.\ R.\ Fletcher, J.\ Phys.\ {\bf C 18}, L55 (1985).

\item \label{bib:barton} G.\ Barton, Ann.\ Phys.\ (N.Y.) {\bf 166}, 322 (1986).

\item \label{bib:sciam} R.\ Y.\ Chiao, P.\ G.\ Kwiat and A.\ M.\
Steinberg, Sc.\ Am.\ {\bf 269}(2), 52 (Aug. 1993).

\end{enumerate}

\end{document}